\documentclass{jfm}
\usepackage{graphicx}
\usepackage{epstopdf, epsfig, color, xcolor}
\usepackage{subfigure}
\usepackage[T1]{fontenc}
\usepackage{newtxtext}
\usepackage[varvw]{newtxmath}

\newcommand{\pard}[2]{\frac{\partial #1}{\partial #2}}

\shorttitle{Global linear stability of a cylinder anchored flame}
\shortauthor{C.-H. Wang, L. Lesshafft and K. Oberleithner}

\title{Global linear stability analysis of a flame anchored to a cylinder}

\author{Chuhan Wang\aff{1,2}\corresp{\email{cwang@ladhyx.polytechnique.fr}}, Lutz Lesshafft\aff{1}
\and Kilian Oberleithner\aff{3}}

\affiliation{\aff{1}LadHyX, CNRS, École Polytechnique, Institut Polytechnique de Paris, 91120 Palaiseau, France 
\aff{2}AML, Department of Engineering Mechanics, Tsinghua University, 100084 Beijing, PR China
\aff{3}Laboratory for Flow Instabilities and Dynamics, Technische Universit\"{a}t Berlin, 10623 Berlin, Germany}

\begin{document}

\maketitle

\begin{abstract}
	This study investigates the linear stability of a laminar premixed
	flame, anchored on a square cylinder and confined inside a channel. Many
	modern linear analysis concepts have been developed and validated around
	non-reacting bluff-body wake flows, and the objective of this paper is
	to explore whether those tools can be applied with the same success to
	the study of reacting flows in similar configurations. It is found that
	linear instability analysis of steady reacting flow states accurately
	predicts critical flow parameters for the onset of limit-cycle
	oscillations, when compared to direct numerical simulation performed
	with a simple one-step reaction scheme in the low Mach number limit.
	Furthermore, the linear analysis predicts a strong stabilising effect of
	flame ignition, consistent with documented experiments and numerical
	simulations. Instability in ignited wake flows is, however, found to set
	in at sufficiently high Reynolds number, and a linear wavemaker analysis
	characterises this instability as being driven by hydrodynamic
	mechanisms of a similar nature as in non-reacting wake flows. The
	frequency of nonlinear limit-cycle flame oscillations in this unstable
	regime is retrieved accurately by linear eigenmode analysis performed on
	the time-averaged mean flow, under the condition that the full set of
	the reacting flow equations is linearised. If, on the contrary,
	unsteadiness in the density and in the reaction rate are excluded from
	the linear model, then the congruence between linear and nonlinear dynamics
	is lost.
\end{abstract}


\section{Introduction}
Bluff-body burners are of particular interest for flame anchoring in a high-velocity flow. The body generates a recirculation zone in its near wake, keeping the flame at a fixed position and preventing blow-off. Flames anchored to bluff bodies have been studied extensively experimentally and numerically \citep{WILLIAMS194821,chen1990comparison,dally1998instantaneous,shanbhogue2009lean,GHANI20154075}. However, vortex shedding may occur in bluff-body wakes, which drastically affects the flame dynamics and may even trigger thermoacoustic instability \citep{manoharan2015absolute,oberleithner2015shear,emerson2016local}. It is therefore important to analyse the physical mechanisms that underpin these oscillations, which can then be targeted for an efficient control. In the field of non-reacting flow, vortex shedding in cylinder wakes is a standard benchmark for stability calculations. The linear stage of oscillation onset in such non-reacting wakes has been rather fully explained within the framework of eigenmode analysis \citep{noack_eckelmann_1994,pier2002frequency,theofilis2003advances,barkley2006linear,taira2020modal}. In the presence of a flame, however, chemical reaction and associated heat release may strongly change the instability dynamics. The objective of this study is to characterise how the presence of a flame in a square cylinder wake, through the effects of chemical reaction and heat release, changes the instability dynamics. To this purpose, a linear eigenmode analysis will be based on the governing equations for a simple reacting flow; the topic of mean flow versus steady flow as base states, as well as ``wavemaker'' concepts, will be revisited in this context.

Linear stability analysis characterises the self-excited behaviour of a flow, by calculating the temporal eigenmodes of the linearised governing equations. Conventionally, the real part of complex eigenvalues denotes the growth rate, and the imaginary part denotes the frequency of flow perturbations. \citet{pier2002frequency} and \citet{barkley2006linear} performed linear stability analysis of the base flow and mean flow of a two-dimensional cylinder wake. Here, the \textit{base} flow refers to a steady solution of the flow equations, and the \textit{mean} flow refers to the temporal average of the oscillating flow state. The least stable eigenmode of the base flow fails to capture the shedding frequency at supercritical Reynolds numbers, whereas the corresponding eigenmode of the mean flow gives excellent predictions. The growth rate of the least stable mean flow eigenmode was found to be almost exactly zero. Similar observations have been made in many other flow cases, prompting the postulation of a ``real-zero-imaginary-frequency'' (RZIF) criterion \citep{turton2015prediction,bengana2021frequency}, which states that perturbations inducing mean flow modifications saturate when the mean flow becomes neutrally stable.  Making use of this criterion, \cite{mantivc2014self} proposed a self-consistent model to predict the frequency and amplitude of vortex shedding without prior knowledge of the mean flow.

With the added complexity of reaction, flame instability may arise from the inherent coupling among various physical mechanisms, including thermoacoustic effects. The classical thermoacoustic instability refers to a scenario where the flame acts as a frequency-selective amplifier of perturbations, and a surrounding structure acts as an acoustic resonator. Recently, a distinct scenario of ``intrinsic thermoacoustic instability'' has also been explored \citep{hoeijmakers2014intrinsic,emmert2015intrinsic,courtine2015dns}, where the acoustic chamber resonance is replaced with an acoustic-hydrodynamic coupling in the upstream system to close the feedback loop. Hydrodynamic instability of a flame, in the literature, may denote at least two distinct concepts: on the one hand, the term is used to refer to instability dynamics that is intrinsic to the flame dynamics itself, without involving any acoustic resonance, such as the Darrieus-Landau and thermo-diffusive instability mechanisms \citep{matalon2007intrinsic}; on the other hand, it is also used to denote instability of the supporting flow field when unsteady reaction dynamics, although present, are not accounted for. Examples for the latter approach include studies of vortex shedding in bluff-body anchored flames \citep{emerson2016local}, shear layer instability \citep{oberleithner2015shear}, the precessing vortex core in swirling flames \citep{oberleithner2015formation}, the coherent structures in turbulent jet flames \citep{kaiser2019prediction,CASEL2022111695}, and inertial waves \citep{albayrak2019propagation,muller2022modal}. The justification of this approach relies on the \textit{a priori} hypothesis that unsteady reaction does not play an active role in the instability mechanism, therefore it is commonly referred to as the ``passive flame'' approach \citep{CASEL2022111695}. 

In reality, however, flame unsteadiness involves a large number of physical processes that may combine to form instability mechanisms. On conceptual grounds, many of these can be either neglected or simplified for a fundamental analysis; in the present context, we choose to regard the low Mach number limit in order to reduce the complexity of compressibility, and we opt for a one-step chemical reaction model that retains much of the specificity of combustion, without overstressing the fuel dependency. 

Bluff bodies of different shapes are used to fix the flame position in various combustion configurations, \textit{e.g.} in propulsion applications, afterburners being one example \citep{shanbhogue2009lean}. The basic scenario is a premixed reactant flow passing around a bluff body, which is ignited in the recirculation region that contains heated reaction products. Fundamental studies of flame dynamics in such a case cover a wide range of topics, such as the near-wall anchoring structures \citep{KEDIA20142327,miguel2016joint}, the onset of flame blow-off \citep{nair2007near,chaudhuri2012blowoff,KEDIA20151304,vance2021investigation,kumar2022blowoff}, the forcing response \citep{shanbhogue2009flame,shin2011dynamics,shin2013flame}, the acoustic flame transfer function \citep{mejia2018influence,kaiser2019impact} and the effect of premixed flow properties, including gas composition \citep{kim2019hydrodynamic,balasubramaniyan2021global} and preheated temperature \citep{erickson2011influence,michaels2017impact}, on flame dynamics and structures. Closely related to our present investigation are studies of the effect of reactant density ratio on the hydrodynamic instability of bluff-body flames by \citep{emerson2012,emerson2015dynamics,suresha2016nonlinear,emerson2016local}. At different preheated reactant temperatures, these authors conducted local stability analysis on experimental velocity and density profiles. A transition from convective to absolute instability was found to occur for preheated lighter unburnt gas when the density ratio between unburnt/burnt gas is close to unity. The authors also estimated the global mode shapes by combining local stability results \citep{juniper2015structural}, for the goal of identifying flow regions related to strong eigenvalue sensitivity \citep{emerson2016local}. The latter is also one of the objectives in the present paper, with an alternative strategy.

We conduct a global linear stability analysis of a two-dimensional premixed laminar flame anchored on a square cylinder. The Navier--Stokes equations for reacting flow are linearised around the steady and mean states, and both the streamwise and cross-stream directions are resolved. Similar approaches have been used in recent years for premixed laminar flames. The closest to our study is the analysis by \citet{avdonin2019thermoacoustic}, who computed the global instability eigenmode of a slot flame in an anechoic chamber, identified as the intrinsic thermoacoustic mode. Those authors also used linearised approaches to calculate flame transfer functions and entropy wave transfer functions \citep{albayrak2018response,meindl2021spurious}. The linear dynamics of M-shaped flames was investigated in a similar framework by \citet{blanchard2015response,blanchard2016pressure}, who focused on the input-output behaviour of the flame, including impulse response and pressure wave generation. Their studies were extended to M-shaped swirling flames by \citet{skene2019adjoint}. In a recent study, we used resolvent analysis to identify the optimal forcing structures in a slot flame, leading to maximal heat release \citep{wang2022linear}. For linear analysis of diffusion jet flames, one may refer to \citet{nichols_schmid_2008}, \citet{qadri2015self}, \citet{moreno-boza_2016}, \citet{qadri2021using} and \citet{sayadi_schmid_2021}. Modal analysis has also been conducted for hot jets \citep{coenen2017global,chakravarthy2018global} and in turbulent swirling flames with the presence of a precessing vortex core \citep{oberleithner2015shear,oberleithner2015formation}.

The reference case of the present study is taken from \citet{KEDIA20142327,KEDIA20151304}, where the authors conducted direct numerical simulations (DNS) of a premixed laminar flame anchored on a solid square cylinder inside a channel at $\mathrm{Re}=500$. The governing equations for those simulations were formulated in the low Mach number limit, and a detailed methane-air reaction scheme of 16 species and 46 steps was implemented. A solid-fluid conjugated heat exchange model was included. \citet{KEDIA20142327} found that the non-reacting cold flow displayed vortex shedding, whereas the ignited reacting flow was steady, anchoring at a position immediately downstream of the square. As one of our objectives is to elucidate the mechanisms by which flame ignition quenches the instability, we use the same geometry and the same flow parameters as \citet{KEDIA20142327}, but we opt for a much simpler chemistry model.

The paper is organised as follows. In \S \ref{sec:methods}, the governing equations and the geometry of the flow configuration are documented. In \S \ref{sec:non-reacting}, nonlinear time stepping and linear analysis of an unstable non-reacting mean flow is carried out. In \S \ref{sec:reacting_base_large}, base flow analysis of the reacting flow is conducted. A detailed wavemaker analysis is attempted in order to identify the spatial region that controls the global flow dynamics. It also gives insight into the important physical mechanisms that contribute to the leading instability eigenmode. In \S \ref{sec:reacting_mean}, we perform nonlinear time stepping of unstable flames at supercritical Reynolds numbers, alongside a modal analysis of the reacting mean flow. Conclusions and perspectives are given in \S \ref{sec:conclusions}.

\section{Methods}
\label{sec:methods}

\subsection{Nonlinear governing equations}

The governing equations for reacting flow are formulated in the low Mach number limit \citep{mcmurtry1986direct}, in terms of primitive variables ($\rho$, $u_x$, $u_y$, $h$, $Y_\mathrm{CH_4}$) in Cartesian coordinates ($x, y$), where ($u_x,u_y$) are the streamwise and cross-stream velocity components, $\rho$ is the density, $h$ is the sensible enthalpy, and $Y_\mathrm{CH_4}$ is the mass fraction of methane. The chemical reaction is modelled by a global one-step scheme for a lean methane-air mixture, requiring only one species equation for $\mathrm{CH_4}$. Each flow variable in the compressible reacting flow equations is expanded in orders of Mach number \citep{mcmurtry1986direct}, and following notations in \citet{albayrak2018response}, these equations are written in the form
\begin{equation}
\label{eq:continu}
\pard{\rho}{t} = -\pard{}{x_j}(\rho u_j),
\end{equation}
\begin{equation}
\label{eq:momentum}
\rho\pard{u_i}{t}= -  \rho u_j\pard{u_i}{x_j}   - \pard{p}{x_i} + \pard{\tau_{ij}}{x_j},
\end{equation}
\begin{equation}
\label{eq:species}
\rho\pard{ Y_\mathrm{CH_4}}{t} = -\rho u_j\pard{Y_\mathrm{CH_4}}{x_j}- \pard{J_j}{x_j} + \dot{\omega}_\mathrm{CH_4},
\end{equation}
\begin{equation}
\label{eq:temperature}
\rho\pard{h}{t} = - \rho u_j\pard{h}{x_j}  - \pard{q_j}{x_j}+ \dot{\omega}_T.
\end{equation}
The system is closed by the equation of state for an ideal gas:
\begin{equation}
\label{eq:perfectgas}
p_0=R_s \rho T.
\end{equation}
The pressure $p_0$ in the equation of state (\ref{eq:perfectgas}) is at the zeroth order of expansion, i.e.~independent of Mach number, whereas $p$ in the momentum equation (\ref{eq:momentum}) is its first-order complement. The molecular stress tensor is given by $\tau_{ij} = -\frac{2\mu}{3}\pard{u_k}{x_k}\delta_{ij} + \mu\left( \pard{u_i}{x_j} + \pard{u_j}{x_i} \right)$, where the molecular viscosity $\mu$ is modelled by the Sutherland law. The flux of species transport and heat transfer are modelled as $J_j=-D_s \pard{Y_\mathrm{CH_4}}{x_j}$ and $q_j=-\alpha\pard{h}{x_j}$, respectively. The transport coefficients therein are associated with a Schmidt number $\mathrm{Sc}=\frac{\mu}{D_s}$ and a Prandtl number $\mathrm{Pr}=\frac{\mu}{\alpha}$. Constant values $\mathrm{Sc}=0.7$ and $\mathrm{Pr}=0.7$ are prescribed in this study, which are valid for a lean methane-air mixture \citep{SATO19821541}. In the equation of state, $R_s$ is the specific gas constant. In the following, the enthalpy is expressed as $h=C_pT$, where the specific heat capacity $C_p$ is taken to be constant. Combining this definition of enthalpy with (\ref{eq:continu}), (\ref{eq:temperature}) and (\ref{eq:perfectgas}), we can eliminate the time derivative term in  (\ref{eq:temperature}) and replace it with
\begin{equation}
\label{eq:control}
0=\rho \pard{u_i}{x_i}+\alpha \rho \pard{}{x_i}\left( \rho^{-2} \pard{\rho}{x_i}\right)+\frac{\rho R_s}{C_p p_0}\dot{\omega}_T,
\end{equation}
in the same way in \citet{kaiser2019prediction}. The system for the following calculations is then given by (\ref{eq:continu})-(\ref{eq:species}) and (\ref{eq:control}).  

The reaction rate is modelled by a one-step chemistry scheme, 
\begin{equation}
\label{eq:1S-chemistry}
\mathrm{CH_4+2O_2\rightarrow CO_2+2H_2O,}
\end{equation}
where the reaction progress rate $\mathcal{Q}$ is given in the form of an Arrhenius law 
\begin{equation}
\mathcal{Q}=[X_\mathrm{CH_{4}}]^{n_{\mathrm{CH_{4}}}}[X_\mathrm{O_{2}}]^{n_{\mathrm{O_{2}}}}T^\beta \exp \left ( -\frac{T_{a}}{T} \right ). \label{eq:arrhenius}\end{equation}

Coefficients for the model constants $n_{\mathrm{CH_{4}}}$, $n_{\mathrm{O_{2}}}$, $\beta$ and $T_a$ are taken from the scheme $\mathrm{1S\_CH4\_MP1}$ provided by \citet{1Step}, which is suitable for a lean premixed methane-air gas. The molar concentrations of $\mathrm{CH_4}$ and $\mathrm{O_2}$ are given by $[X_{\mathrm{CH_4}}]=\rho \frac{Y_{\mathrm{CH_4}}}{W_{\mathrm{CH_4}}}$ and $[X_{\mathrm{O_2}}]=\rho \frac{Y_{\mathrm{O_2}}}{W_{\mathrm{O_2}}}$, respectively, where $W_\mathrm{CH_4}$ and $W_\mathrm{O_2}$ denote their molecular masses. The reaction rate in the species equation is given by $\dot{\omega}_\mathrm{CH_4}=-W_{\mathrm{CH_4}}\mathcal{Q}$, and the heat release due to combustion in the enthalpy equation is given by $\dot{\omega}_T=-\Delta h_{f}^o \mathcal{Q}$, where $\Delta h_{f}^o$ is the standard enthalpy of reaction.

The reacting flow model documented in this section relies on several simplifying assumptions; in particular, the one-step reaction scheme (\ref{eq:1S-chemistry})-(\ref{eq:arrhenius}) does not reflect the rich variety of chemical reactions that take place in methane combustion. It is chosen here for convenience, mainly to maintain a small number of empirical model parameters and species to track for this fundamentally oriented study. As will be shown in \S{}\ref{sec:reacting_base}, this model seems to appropriately reproduce the steady states obtained numerically by \cite{KEDIA20142327} with more detailed reaction kinetics.
It should be understood that all results presented in this paper, and the conclusions inferred therefrom, can be valid only within the limitations of the flow model.

\subsection{Calculation of base flow and mean flow}
\begin{figure} 
	\centering\includegraphics[width=0.5\textwidth]{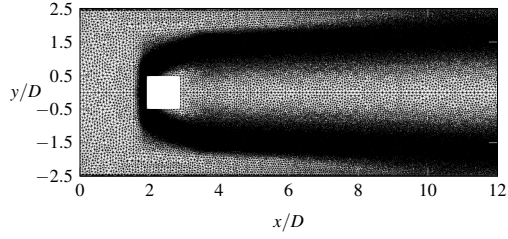}
	\caption{Unstructured mesh of the numerical domain. Lengths are normalised with the square width $D= 5$ mm.} 
	\label{fig:para_mesh}
\end{figure}

The geometry of the numerical domain, presented in figure~\ref{fig:para_mesh}, corresponds to the simulations of \citet{KEDIA20142327,KEDIA20151304} with a square cylinder of side length $D=5$ mm centred in a channel that is 25 mm wide. The numerical domain extends $2D$ upstream and $9D$ downstream of the cylinder. Inflow conditions for streamwise velocity and temperature are set as $U=1.6$ m/s and $T_0=300$ K, which leads to an inflow Reynolds number $\mathrm{Re} = \rho_0 U D/\mu_0 = 500$, where $\rho_0$ and $\mu_0$ denote the density and the molecular viscosity at the inflow, respectively. The inlet velocity profile is prescribed as being parabolic, and a no-slip adiabatic condition is used at the channel walls. A no-slip condition is also prescribed on the cylinder surface, but an important difference with respect to the reference simulations is that we prescribe a constant temperature of the cylinder, $T_c$. A no-flux condition of species transport is imposed on the cylinder and channel walls. \citet{KEDIA20142327} studied the equivalence ratio $\phi$ in the range from 0.5 to 0.7. We revisit this range and consider leaner mixtures down to $\phi=0.25$.

The nonlinear governing equations are discretised on the unstructured mesh shown in figure \ref{fig:para_mesh} with a continuous Galerkin method, as provided by the FEniCS software \citep{alnaes2015fenics}. The mesh contains 41\,145 triangular cells, with high spatial resolution in the flame front region.

A steady flow solution is obtained by Newton's iterative method; such a steady state will be denoted a ``base flow'' in the following. Convergence of Newton's method is conditioned on the initial guess, which must be sufficiently close to the solution. In particular, it is difficult to displace a sharp flame front during the iterations, but much less difficult to successively steepen an initially diffuse front. An adapted approach, which was used to calculate the steady solution of slot flames \citep{albayrak2018response,douglas2021dynamics}, is applied in the present case. All diffusion coefficients are multiplied by an artificial parameter $c_1$, and the heat release term is multiplied by a parameter $c_2$. We first enlarge the convergence radius by computing a very viscous non-reacting flow as the initial guess. To this end, $c_1$ is set to 30, and $c_2$ is set to 0. Then we gradually decrease $c_1$ from 30 to 1, and at the same time we increase $c_2$ from 0 to 1. Each time $c_1$ and $c_2$ are changed, one Newton iteration is performed, until both factors are brought to 1. The final base flow is thus recovered with residual errors of the order of machine precision, after around 25 iteration steps.

To obtain the time-averaged flow fields in unstable configurations, the nonlinear reacting flow equations are integrated in time with a Crank-Nicolson scheme. Averaging is started when the periodic regime is reached. Such time-averaged fields will be called the ``mean flow'' throughout this study.

\subsection{Global linear analysis}
Flow fluctuations $\boldsymbol{q'}(\boldsymbol{x},t)$ around the steady base state or around the time-averaged mean state are assumed to be governed by the linear equation
\begin{equation}
\mathsfbi{B}\pard{\boldsymbol{q'}}{t}	=\mathsfbi{A}\boldsymbol{q'},
\end{equation}
where the matrices $\mathsfbi{B}$ and $\mathsfbi{A}$ are built from linearisation of (\ref{eq:continu})-(\ref{eq:species}) and (\ref{eq:control}). This linear assumption is valid for infinitesimally small fluctuations around the base state, and it is hoped to be extendable to finite-amplitude fluctuations around the mean flow in supercritical conditions \citep{barkley2006linear,mantivc2014self}. The linearisation of the governing equations is performed symbolically within FEniCS by use of the Unified Form Language \citep{FENICS_UFL}. The flow fluctuations $\boldsymbol{q'}(\boldsymbol{x},t)$ can be expanded in the basis of eigenmodes $\boldsymbol{q}'_j(\boldsymbol{x},t)=\boldsymbol{\phi}_j(\boldsymbol{x})\exp(\lambda_j t)$ obtained from the generalised eigenvalue problem
\begin{equation}
\label{eq:direct}
\lambda_j \mathsfbi{B}\boldsymbol{\phi}_j	=\mathsfbi{A}\boldsymbol{\phi}_j.
\end{equation}
The eigenvalues $\lambda_j$ and associated eigenvectors $\boldsymbol{\phi}_j$ are computed with a Krylov method. We define the frequency as $2\pi f_j=-Im[\lambda_j]$ and the growth rate as $2\pi \sigma_j=Re[\lambda_j]$, so that $\lambda_j/(2\pi)=\sigma_j-i f_j$. In the following, the eigenmode with maximal growth rate will be named the \textit{leading} eigenmode, with eigenvalue $\lambda_0$ and eigenvector $\boldsymbol{\phi}_0$. 

Homogeneous Dirichlet boundary conditions are prescribed for all fluctuations, except pressure, at the inflow; no-slip adiabatic conditions are imposed at the lateral channel walls; and no-slip isothermal conditions, i.e.~zero fluctuations of density and velocity, are chosen on the cylinder surface. Stress-free boundary conditions at the outflow are used, which is a common and convenient choice in finite-element formulations for low spurious reflections \citep[e.g.][]{lesshafft2018artificial}.

The mesh convergence is tested by using $74\,850$, $91\,776$ and $155\,636$ cells to calculate the leading eigenmode at $\phi=0.5$ for the ignited reacting flow. The relative error of the leading eigenvalue is less than 0.85\%.

\section{Non-reacting flow}
\label{sec:non-reacting}
\begin{figure} 
	\centering\includegraphics[width=0.5\textwidth]{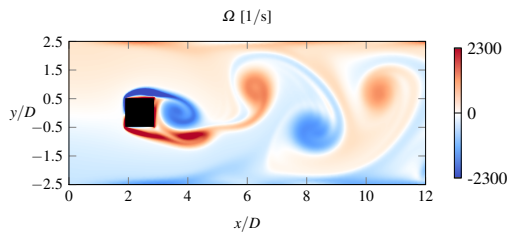}
	\caption{Snapshot of the vorticity field of the non-reacting flow, obtained by nonlinear time stepping.}
	\label{fig:para_non_reacting_vorticity}
\end{figure}

Nonlinear simulation is first performed for a non-reacting flow at $\mathrm{Re}=500$. The terms related to chemical reaction, $\dot{\omega}_\mathrm{CH_4}$ and $\dot{\omega}_T$, are removed from the governing equations for this calculation, and the cylinder temperature $T_c$ is set to be the same as the inflow temperature. Figure \ref{fig:para_non_reacting_vorticity} shows the periodic regime of vortex shedding, in agreement with figure 4 of \cite{KEDIA20142327}. Note that due to the parabolic inlet profile and the confinement by no-slip channel walls, the vortex dynamics observed here is different from an unbounded uniform flow past a square cylinder \citep{davis1984numerical,suzuki1993unsteady,turki2003two}.

\begin{figure} 
	\centering\includegraphics[width=0.8\textwidth]{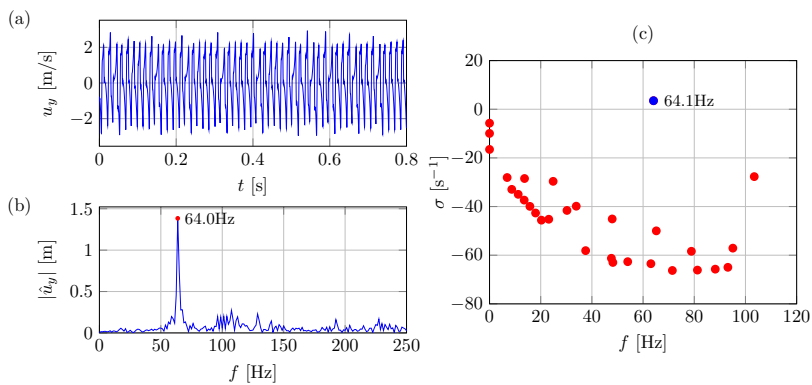}
	\caption{(a) Time evolution of $u_y$ measured at location $(x/D,y/D)=(4.5, 0)$ in the non-reacting flow. (b) Frequency spectrum of the signal in (a). (c) Eigenvalues of the non-reacting mean flow. The unstable eigenvalue has $f=64.1$ Hz and $\sigma=3.5$ $\mathrm{s^{-1}}$, marked in blue.}
	\label{fig:xy_non_reacting_mean}
\end{figure}

The temporal signal of the cross-stream velocity $u_y$, measured at the location $(x/D,y/D)=(4.5, 0)$, is shown in figure \ref{fig:xy_non_reacting_mean}\textit{a}, and its frequency spectrum is given in figure \ref{fig:xy_non_reacting_mean}\textit{b}. The temporal signal is quasi-periodic and appears to be near the onset of chaotic dynamics. A distinct peak at 64.0 Hz corresponds to the vortex shedding frequency.  

The nonlinear flow fields are averaged over the time horizon shown in figure~\ref{fig:xy_non_reacting_mean}\textit{a}, and perturbation eigenmodes are computed for this mean flow. A leading eigenvalue at 64.1 Hz is identified, in excellent agreement with the measured vortex shedding frequency. Here, the growth rate is found to be $\sigma=3.5$ $\mathrm{s^{-1}}$, relatively close to zero, therefore fairly consistent with the RZIF criterion. This slightly non-zero growth rate may be attributed to the weak non-periodicity of the unsteady flow \citep{mantivc2014self}. More details about this non-reacting flow analysis results are given in the Appendix, including a comparison of eigenmodes obtained for base flow and mean flow, and convergence tests with respect to the domain length and to the time-averaging interval used in the construction of the mean flow.       

\section{Reacting flow: eigenmodes of the steady base state}
\label{sec:reacting_base_large}
\subsection{Calculation of the base flow}
\label{sec:reacting_base}
\begin{figure} 
	\centering\includegraphics[width=0.8\textwidth]{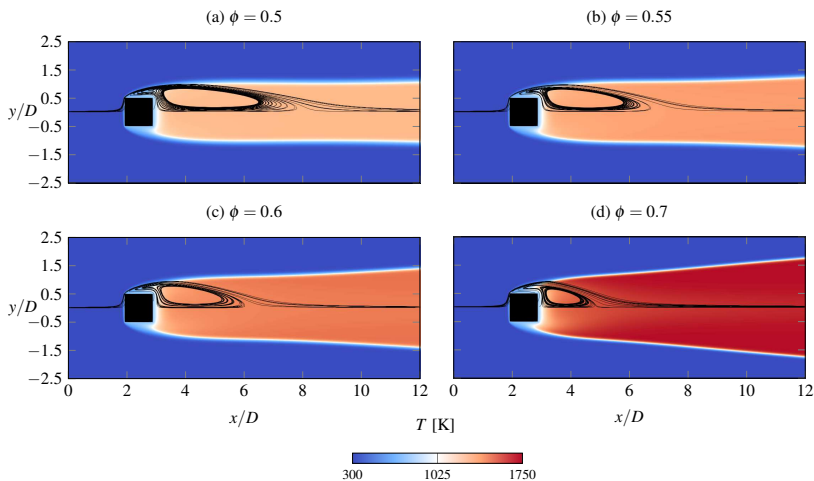}
	\caption{Reacting base flow at equivalence ratio $\phi$ from 0.5 to 0.7. Streamlines are superposed on the upper half of the temperature fields. The temperature of the cylinder wall is fixed at 750 K. }
	\label{fig:para_kedia}
\end{figure}

\begin{figure} 
	\centering\includegraphics[width=0.4\textwidth]{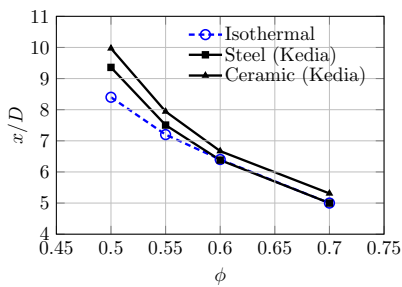}
	\caption{The end position of the recirculation bubble at $y=0$ calculated with an isothermal cylinder wall at $T_c=750$ K, compared with figure 8 of \cite{KEDIA20142327} using conjugate heat exchange for steel and ceramic.  }
	\label{fig:xy_kedia}
\end{figure}

The steady base state of a reacting flow is computed at various values of equivalence ratio $\phi$. The Reynolds number is always 500, and the inflow temperature is maintained at 300 K. The cylinder temperature $T_c$ is always fixed at $T_c=750$ K, a value close to the temperature profiles in figure 13 of \cite{KEDIA20142327}. Such a condition is consistent with the reference, where the authors stated that the cylinder was almost isothermal for high conductivity cases. The steady base states of the reacting flow at $0.5\leq\phi\leq 0.7$ are shown in figure \ref{fig:para_kedia}, in the same range of equivalence ratio investigated in \cite{KEDIA20142327}. The structure of the computed base flows match well with figure 7 of \cite{KEDIA20142327}, in spite of our different temperature condition at the cylinder and the simplified reaction scheme. The resulting Damköhler number is 33.3 at $\phi=0.7$, and 2.38 at $\phi=0.5$. The flame front is anchored on the cylinder, advancing towards the channel walls with downstream distance. A recirculation zone, immediately behind the cylinder, is spatially separated from the flame fronts. Figure \ref{fig:xy_kedia} shows that the length of the recirculation bubble grows with the reduction of the equivalence ratio, qualitatively in agreement with figure 8 of \cite{KEDIA20142327}. As leaner mixtures will be considered, in the following we use a longer computational domain extending downstream to $x/D=20$ such that the entire recirculation bubble is included. 

\subsection{Stabilisation by ignition: a wavemaker analysis}
\begin{figure} 
	%\centering\includegraphics[width=0.4\textwidth]{figures/para_kedia/xy_phi_growth-figure0.pdf}
	\centering\includegraphics[width=0.4\textwidth]{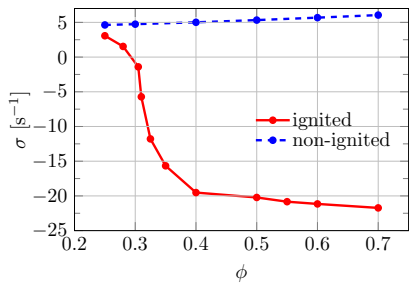}
	\caption{Growth rates $\sigma$ of the least stable base flow eigenmode as a function of equivalence ratio $\phi$ in the ignited (red) and non-ignited (blue) cases. }
	\label{fig:xy_phi_growth}
\end{figure}

The steady states of the flame computed are obtained by Newton iteration, and it is not known \textit{a priori} if they are linearly stable. Therefore, eigenmodes of the steady states are calculated, and the growth rate of the least stable eigenvalue is plotted in figure \ref{fig:xy_phi_growth} as a red line. Over the range of equivalence ratios $0.5 \le \phi \le 0.7$, where \cite{KEDIA20142327} report steady dynamics from their simulations, the flame is indeed found to be stable in our calculations, as characterised by a negative maximum growth rate.

\begin{figure} 
	\centering\includegraphics[width=\textwidth]{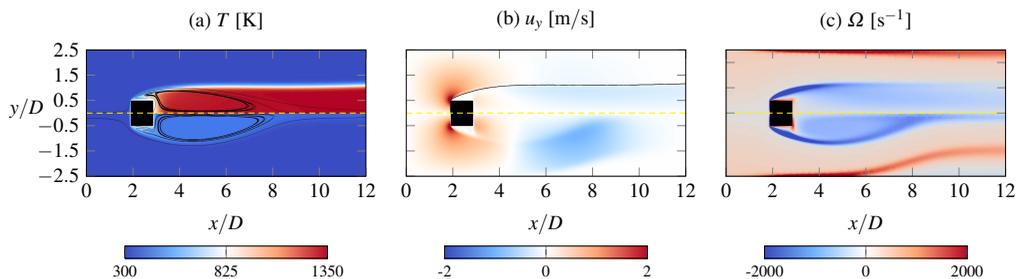}
	\caption{Comparison of base flow quantities between ignited (upper half, $\sigma=5.3~\mathrm{s^{-1}}$) and non-ignited case (lower half, $\sigma=-20.3~\mathrm{s^{-1}}$) at $\phi=0.5$. Streamlines are superposed on the temperature fields in (a). Temperature isocontours of $T=550$ K are superposed on cross-stream velocity in (b), indicating the flame front position. Vorticity fields are shown in (\textit{c}).}
	\label{fig:para_ignited}
\end{figure}

Aiming at understanding the stabilising effect by ignition, a non-ignited flow is calculated at different equivalence ratios using the same value of cylinder temperature $T_c=750$ K. The non-ignited configurations are also obtained through Newton's method, but using the cold flow with $T_c=300$ K as initial guess of iteration. The steady states at various $T_c$ with an increment of 50 K are calculated consecutively up to $T_c=750$ K. In this way, the flow field is always non-ignited. The ignited and the non-ignited flow fields at $\phi=0.5$ are shown in the upper and lower halves of figure \ref{fig:para_ignited}, respectively. The non-ignited flow field closely resembles the non-reacting flow around a hot cylinder. Again, both states are steady solutions of the reacting flow equations with exactly the same flow parameters. The growth rate of the leading mode of the non-ignited flow at various equivalence ratios is shown as a blue line in figure \ref{fig:xy_phi_growth}. The non-ignited flow is always found to be unstable with a nearly constant positive growth rate. At $\phi=0.5$, the growth rate at non-ignited conditions is 5.3 $\mathrm{s^{-1}}$, while the ignited flow is stable with the largest growth rate of -20.3 $\mathrm{s^{-1}}$. Such a stabilising effect of ignition has often been observed in experiments and numerical simulations \citep{bill1986effect,mehta2003combustion,erickson2011influence,KEDIA20142327,oberleithner2015formation,GEIKIE202128}.
%\centering\includegraphics[width=\textwidth]{figures/para_kedia/para_ignited-figure0.pdf}

We now attempt to characterise the cause of this stabilisation from a ``wavemaker'' analysis of the leading eigenmode. Such an analysis aims at identifying flow regions where local feedback mechanisms act to give rise to intrinsic oscillations. These flow regions are characterised by the sensitivity of the eigenvalue to structural changes in the linear operator \citep{giannetti2007structural}.

First, the \emph{adjoint eigenvectors} $\boldsymbol{\phi}_j^\dag$ are introduced as the eigenvectors of the conjugate-transposed system matrices of (\ref{eq:direct}):
\begin{equation}
    \label{eq:adjoint}
    \lambda_j^* \mathsfbi{B}^H\boldsymbol{\phi}_j^\dag	=\mathsfbi{A}^H\boldsymbol{\phi}_j^\dag.
\end{equation}

A \emph{direct eigenvector} $\boldsymbol{\phi}_j$ and an adjoint eigenvector $\boldsymbol{\phi}_j^\dag$ are thus associated by their complex conjugated eigenvalues. The sets of direct and adjoint eigenvectors, normalised appropriately, fulfill the biorthogonality relation \citep{hill1995adjoint},
\begin{equation}
    \label{eq:biorthogonal}
	\langle \boldsymbol{\phi}_j^\dag,\mathsfbi{B}\boldsymbol{\phi}_k\rangle = \delta_{jk},
\end{equation}
where the spatial inner product over the flow domain $\Omega$ is defined as
\begin{equation}
	\langle \boldsymbol{g}(\boldsymbol{x}),\boldsymbol{h}(\boldsymbol{x})\rangle=\int_\Omega \boldsymbol{g}^* \boldsymbol{h} d\boldsymbol{x}.
\end{equation}

The wavemaker, as identified by the most commonly used definition of \cite{giannetti2007structural}, is the local Frobenius norm of the structural sensitivity tensor, which is constructed as a local product of the direct and associated adjoint eigenmode. Flow regions where this quantity is large are expected to contribute strongly to the global flow instability.
For flames anchored to a cylinder, such a wavemaker analysis was carried out by \citet{emerson2016local}. Global eigenmodes were estimated by combining results from local analysis \citep{juniper2015structural}, and the wavemaker region was found to lie in the recirculation zone of the wake. In the analysis of \citet{emerson2016local}, it was assumed \textit{a priori} that the instability is shear-driven: only the velocity and density fields were taken as input, whereas the effects of unsteady chemical reaction were not taken into account. 

In this study, we attempt a wavemaker analysis without the quasi-parallel assumption, and we include perturbations of species concentration and enthalpy. A variation of the wavemaker definition is adopted in the following, which takes into account the entire state vector and allows an evaluation of contributions from individual components of the linear operator. This method was proposed by \cite{Marquet2015Identifying}, and applied for instance by \cite{chakravarthy2018global} and \cite{murali2022stability}. From (\ref{eq:direct}) and (\ref{eq:biorthogonal}), the leading eigenvalue is obtained in the form of a spatial integral:
\begin{equation}
\label{eq:endo}
	\lambda_0=\langle \boldsymbol{\phi}_0^\dag,\lambda_0\mathsfbi{B}\boldsymbol{\phi}_0 \rangle=\langle \boldsymbol{\phi}_0^\dag,\mathsfbi{A}\boldsymbol{\phi}_0 \rangle=\int_\Omega \boldsymbol{E}(\boldsymbol{x}) d\boldsymbol{x}.
\end{equation}
The integrand $\boldsymbol{E}(\boldsymbol{x})$ is given by the spatially local product between $\boldsymbol{\phi}_0^\dag$ and $\mathsfbi{A}\boldsymbol{\phi}_0$. \cite{Marquet2015Identifying} name $\boldsymbol{E}(\boldsymbol{x})$ the ``endogeneity'' of the eigenmode, and we will adopt this word for ease of writing. The growth rate $Re(\lambda_0)/(2\pi)$ is given by the spatial integral of $Re(\boldsymbol{E})/(2\pi)$, therefore regions where $Re(\boldsymbol{E})$ is positive are expected to contribute to the destabilisation of the leading eigenmode; conversely, flow regions with negative values are interpreted as having a stabilising influence.

\begin{figure} 
	\centering\includegraphics[width=\textwidth]{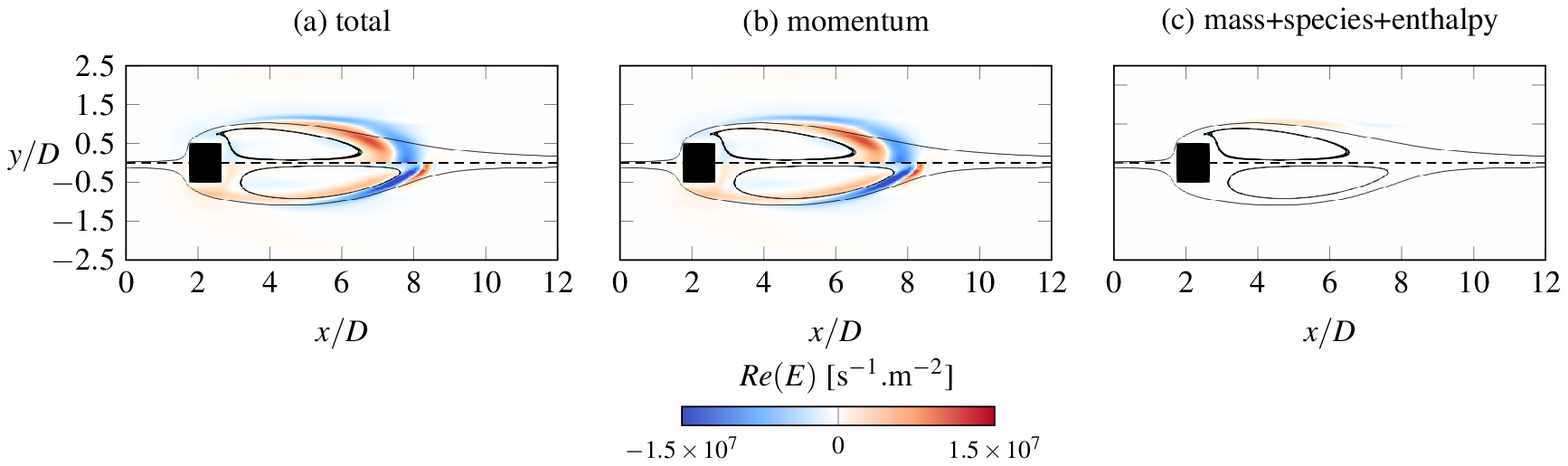}
	\caption{Endogeneity field $Re\left[\boldsymbol{E}(\boldsymbol{x})\right]/(2\pi)$ corresponding to the least stable eigenmode in the ignited and non-ignited cases at $\phi=0.5$: wavemaker according to the definition by \cite{Marquet2015Identifying}. (a) Full endogeneity, with spatial integral $Re(\lambda_0)/(2\pi)=5.3$ $\mathrm{s^{-1}}$ for the ignited case and $Re(\lambda_0)/(2\pi)=-20.3$ $\mathrm{s^{-1}}$ for the non-ignited case. (b) Contribution of the momentum perturbation. (c) Contribution of the sum of the other three linearised equations. The streamlines are superposed. Top of each plot: ignited. Bottom of each plot: non-ignited.}
	\label{fig:para_endo}
\end{figure}

Figure \ref{fig:para_endo}\textit{a} displays $Re(\boldsymbol{E})/(2\pi)$ for the ignited (top) and non-ignited (bottom) flow shown in figure \ref{fig:para_ignited}. For both cases, the instability seems to be generated mainly in the shear region of the recirculation bubble in the near wake, close to the downstream stagnation point. This wavemaker region in figure \ref{fig:para_endo}\textit{a} closely resembles the non-reacting cylinder wake results of \cite{giannetti2007structural}, indicating that similar instability mechanisms are at play. The dominant feature in the endogeneity plot is an inner positive region and an outer negative region appearing at the shear layer near the downstream end of the recirculation bubble, identified as the flow region that most strongly influences the instability growth rate.

%\centering\includegraphics[width=\textwidth]{figures/para_kedia/para_endo-figure0.pdf}

The matrix $\mathsfbi{A}$ in the wavemaker definition (\ref{eq:endo}) can be split into components that represent individual equations or terms of the linearised flow equations, and the isolated contributions of these equations or terms can be visualised as partial endogeneity fields \citep{Marquet2015Identifying,chakravarthy2018global}. To study quantitatively the contribution to the global instability from different physical mechanisms, we decompose $\mathsfbi{A}$ into four matrices in the form
\begin{equation}
	\mathsfbi{A}=\mathsfbi{A}_{\mathrm{mass}}+\mathsfbi{A}_{\mathrm{momentum}}+\mathsfbi{A}_{\mathrm{species}}+\mathsfbi{A}_{\mathrm{enthalpy}},
\end{equation}
where $\mathsfbi{A}_{\mathrm{mass}}$, $\mathsfbi{A}_{\mathrm{momentum}}$, $\mathsfbi{A}_{\mathrm{species}}$ and $\mathsfbi{A}_{\mathrm{enthalpy}}$ are the linear operators corresponding to the conservation equations of mass, momentum, species and enthalpy in matrix $\mathsfbi{A}$. Figure \ref{fig:para_endo}\textit{a} shows the total endogeneity field associated with $\mathsfbi{A}$. The partial endogeneity field of momentum equations ($\mathsfbi{A}_{\mathrm{momentum}}$) is given in figure \ref{fig:para_endo}\textit{b}, and that representing the sum of mass, enthalpy and species equations ($\mathsfbi{A}_{\mathrm{mass}}+\mathsfbi{A}_{\mathrm{species}}+\mathsfbi{A}_{\mathrm{enthalpy}}$) is given in figure \ref{fig:para_endo}\textit{c}. A comparison of figures \ref{fig:para_endo}\textit{a} and \ref{fig:para_endo}\textit{b} clearly shows that the dominant contributions to the growth rate stem from the linearised momentum equation, with nearly identical distributions of the endogeneity field between $\mathsfbi{A}$ and $\mathsfbi{A}_{\mathrm{momentum}}$. Figure \ref{fig:para_endo}\textit{c} reveals that the combined effect of the other three linearised conservation equations on the growth rate is weak: only a small positive region around the flame front in the near wake emerges in the endogeneity field for the ignited case. The corresponding endogeneity of the non-ignited case can hardly be observed with the same range of colour bar, consistent with the fact that barely any reaction takes place. A more detailed analysis (not shown) reveals that contributions to the growth rate are dominated by competition between the destabilising perturbation production term $\rho(\boldsymbol{u}'\cdot\nabla)\boldsymbol{U}$ and the stabilising convection term $\rho(\boldsymbol{U}\cdot\nabla)\boldsymbol{u}'$ in the linearised momentum equation. The stabilising contribution from momentum diffusion is small in comparison.

We now examine why the presence of a flame stabilises the flow. Figures \ref{fig:para_ignited}\textit{a-c} show the temperature, cross-stream velocity and vorticity fields of the ignited (upper half) and non-ignited (lower half) base flow, respectively. The downstream end of the recirculation bubble is where the wavemaker is located, and that region looks very different in the non-ignited and ignited cases. The non-ignited flow closely resembles a standard cylinder flow where there is strong shear near the downstream stagnation point. In the ignited case, the thermal expansion has a significant effect on the base flow, where much less shear is observed at the end of the recirculation bubble, as revealed by the vorticity field in figure \ref{fig:para_ignited}\textit{c}.    

We find that the reaction acts in three ways in modifying the base flow and its stability. First, the recirculation bubble is quenched because of the thermal expansion effect. Second, the ignited flame region of around 1350 K in the present case increases the kinematic viscosity $\mu$ by a factor of three according to the Sutherland law, but also the kinematic viscosity $\nu=\frac{\mu}{\rho}$ by a factor of ten due to the decrease of density, with respect to the cold flow region of 300 K. The significantly higher viscosity potentially stabilises the perturbations and diffuses the shear layer. Third, there is a strong pressure drop behind the obstacle in the non-ignited case, while in the ignited case, this pressure deficit is filled by the expanding fluid. As shown in figure \ref{fig:para_ignited}\textit{b}, the flow is pulled strongly towards the centreline at $6<x/D<8.5$ in the non-ignited case, while such an intense movement is not observed in the ignited case, resulting in much less shear in the wake.

\subsection{Effect of equivalence ratio}
\begin{figure} 
	\centering\includegraphics[width=\textwidth]{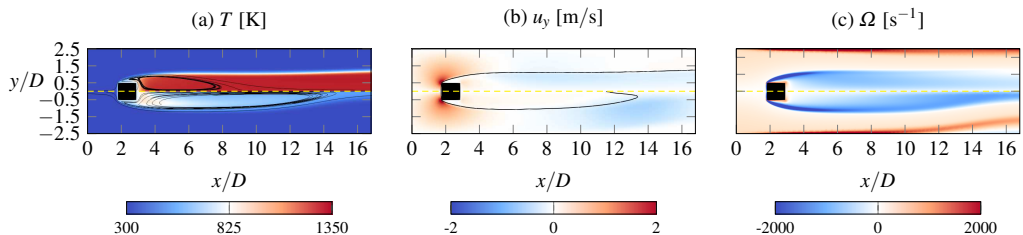}
	\caption{Comparison of base flow quantities between ignited flame at $\phi=0.5$ (upper half, $\sigma=3.1~\mathrm{s^{-1}}$) and at $\phi=0.25$ (lower half, $\sigma=-20.3~\mathrm{s^{-1}}$). Streamlines are superposed on the temperature fields in (a). Temperature isocontours of $T=550$ K are superposed on cross-stream velocity in (b). Vorticity fields are shown in (\textit{c}).}
	\label{fig:para_phi}
\end{figure}

We use the same argument as in the previous subsection to explain the effect of equivalence ratio on the global instability. While the lean flames considered here are still fully ignited, reducing the equivalence ratio continuously decreases the effect of reaction. Figure \ref{fig:xy_phi_growth} shows that decreasing the equivalence ratio $\phi$ destabilises the flame, leading to global instability at $\phi<0.3$. Such a result is consistent with several experimental studies confirming that vortex shedding is observed only at lean conditions \citep{hertzberg1991vortex,kiel2007detailed,nair2007near,balasubramaniyan2021global}. Figures \ref{fig:para_phi}\textit{a-c} show the base flow quantities of the ignited flames at $\phi=0.5$ (upper half) and $\phi=0.25$ (lower half). By reducing the equivalence ratio, the region of the recirculation bubble is extended and the temperature-dependant viscosity decreases further. Consequently, a stronger shear at the end of the recirculation bubble is observed at $\phi=0.25$, shown in figure \ref{fig:para_phi}\textit{c}, eventually leading to a positive growth rate ($\sigma=3.1~\mathrm{s^{-1}}$).

%\centering\includegraphics[width=\textwidth]{figures/para_kedia/para_phi-figure0.pdf}

\section{Reacting flow: eigenmodes of the mean state}
\label{sec:reacting_mean}
\subsection{Unstable flames at supercritical Reynolds numbers}

\begin{figure} 
	\centering\includegraphics[width=0.99\textwidth]{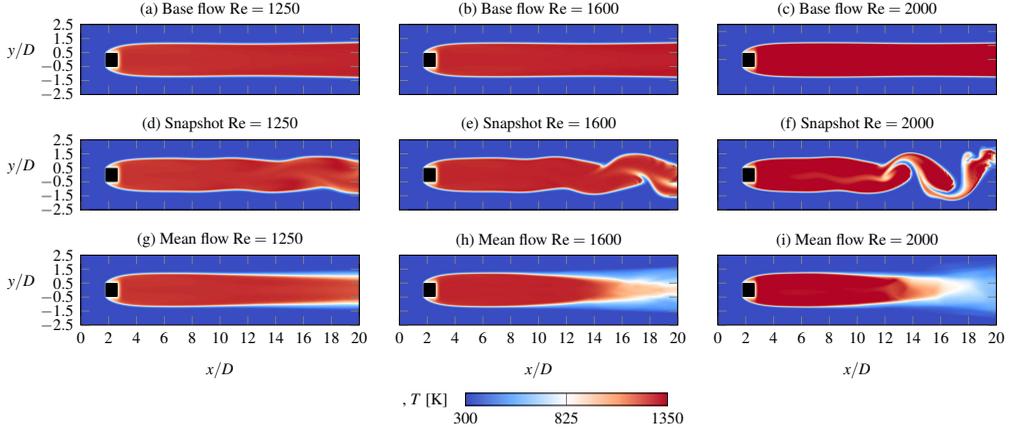}
	\caption{Snapshots, mean flow and base flow of the reacting flow at three supercritical Reynolds numbers with $T_c=750$ K and $\phi=0.5$. Temperature fields are shown.}
	\label{fig:para_mean_T}
\end{figure}

\begin{figure} 
	\centering\includegraphics[width=0.8\textwidth]{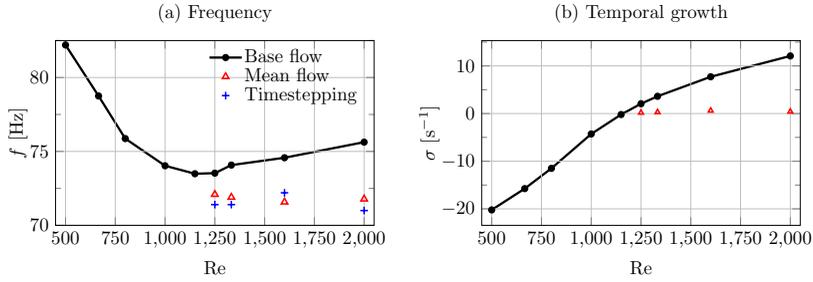}
	\caption{Frequency $f$ and growth rates $\sigma$ of the leading eigenmode in the base flow (black line) and mean flow (red triangles) as a function of $\mathrm{Re}$, at $\phi=0.5$. The oscillation frequencies obtained through time stepping are represented as blue crosses.}
	\label{fig:xy_Re}
\end{figure}

We increase the Reynolds number and investigate the unsteady ignited flow behaviour at $\phi=0.5$. Eigenmode analysis is conducted on each base flow obtained in the range from $\mathrm{Re}=500$ to $\mathrm{Re}=2000$, of which some are presented in figures \ref{fig:para_mean_T}\textit{a}-\textit{c}. The frequency and growth of leading eigenvalues are presented in figure \ref{fig:xy_Re}. At $\mathrm{Re}<1150$, the leading mode has negative growth and the flow therefore is stable. The flow becomes unstable at higher Reynolds numbers. 
%\centering\includegraphics[width=0.99\textwidth]{figures/para_kedia/para_mean_T-figure0.pdf}

%\centering\includegraphics[width=0.8\textwidth]{figures/para_kedia/xy_Re-figure0.pdf}

The temporal snapshots shown in figures \ref{fig:para_mean_T}\textit{d}-\textit{f} at $\mathrm{Re}=1250$, 1600 and 2000, are produced using nonlinear time stepping. Antisymmetric oscillations are observed in all cases, and the amplitude of oscillation increases with $\mathrm{Re}$. The temporal signal of cross-stream velocity $u_y$ at the location $(x/D,y/D)=(8, 0)$ is used to calculate the frequency of vortex shedding, where the oscillation is observed in the regime of limit cycle. The flame further downstream displays quasi-chaotic behaviours at $\mathrm{Re}=2000$. The oscillation frequencies measured are represented as blue crosses in figure \ref{fig:xy_Re}\textit{a}, revealing that the oscillation frequencies are almost constant in the range of Reynolds numbers considered.

\begin{figure} 
	\centering\includegraphics[width=0.8\textwidth]{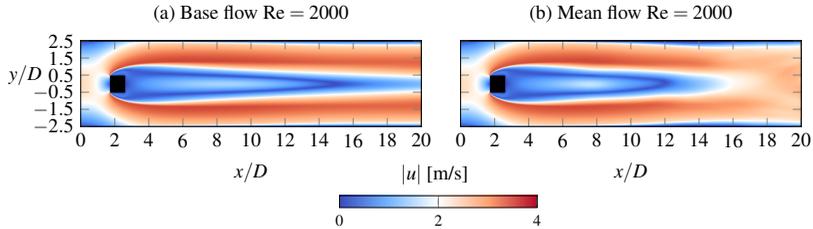}
	\caption{Base flow and mean flow of velocity magnitude at $\mathrm{Re}=2000$ with $T_c=750$ K and $\phi=0.5$.}
	\label{fig:para_mean_u}
\end{figure}

The time-averaged mean flow is represented by the temperature field in figures \ref{fig:para_mean_T}\textit{g}-\textit{i}. The flame front becomes significantly thicker downstream at $\mathrm{Re}=1250$. Furthermore, the mean flow is largely distorted downstream at $\mathrm{Re}=1600$ and $2000$, leading to a shortened flame length. An apparently shorter wake is also observed in the mean flow at $\mathrm{Re}=2000$ in comparison with its base state counterpart, shown in figure \ref{fig:para_mean_u}.
%\centering\includegraphics[width=0.8\textwidth]{figures/para_kedia/para_mean_u-figure0.pdf}

The leading eigenvalues of the mean flow are represented as red triangles in figure \ref{fig:xy_Re}. Their frequencies accurately match those measured in the nonlinear simulations (blue crosses), with a relative error of less than 1\%. In contrast, the frequencies predicted from the base flow analysis deviate with increasing distance from the critical Reynolds number. The base flow eigenmode is unstable above $\mathrm{Re} = 1150$, while the mean flow mode has a growth rate of nearly zero for all supercritical conditions, which is consistent with the classical results for non-reacting cylinder wakes \citep{barkley2006linear} and the RZIF criterion \citep{turton2015prediction,bengana2021frequency}.

\begin{figure} 
	\centering\includegraphics[width=\textwidth]{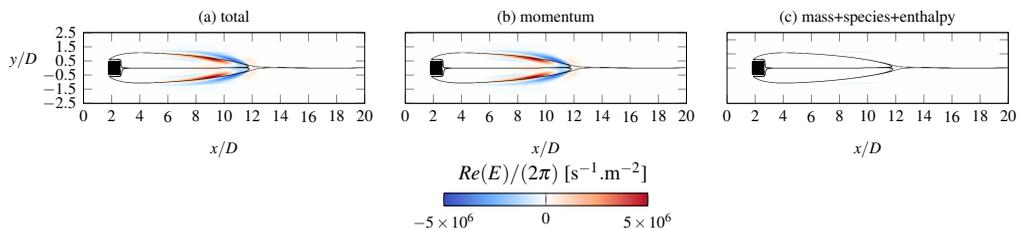}
	\caption{Endogeneity field $Re\left[\boldsymbol{E}(\boldsymbol{x})\right]/(2\pi)$ corresponding to the least stable mean flow eigenmode of unsteady ignited flame at $\mathrm{Re}=1250$: wavemaker according to the definition by \cite{Marquet2015Identifying}. (a) Full endogeneity, with spatial integral $Re(\lambda_0)/(2\pi)=0.2$ $\mathrm{s^{-1}}$. (b) Contribution of the momentum perturbation. (c) Contribution of the sum of the other three linearised equations. The streamlines are superposed.}
	\label{fig:para_mean_endo}
\end{figure}

The wavemaker of the leading mean flow eigenmode, at $\mathrm{Re}=1250$, is displayed in figure \ref{fig:para_mean_endo} in the same way as for the base flow eigenmode in figure \ref{fig:para_endo}. Again, it is seen that the growth rate of the mean flow eigenmode is controlled by terms in the momentum equation alone, and the zero net growth of this saturated mode arises from the balance of stabilising and destabilising dynamics in the downstream part of the recirculation bubble. Species and enthalpy equations, which contain the reaction terms, have virtually no influence on the growth rate. This result again suggests that the saturated instability dynamics of the finite-amplitude oscillations are underpinned by ``hydrodynamic'' effects, as opposed to reaction-driven mechanisms like flame wrinkling. This conclusion is, of course, limited to the present flow case.

The question now arises whether species and enthalpy fluctuations are irrelevant to the instability dynamics, to the point that they can be removed altogether from the linear model. Such reduced ``passive flame'' models have been tried, often successfully, in recent literature \citep[e.g.][]{CASEL2022111695}.

%\centering\includegraphics[width=\textwidth]{figures/para_kedia/para_mean_endo-figure0.pdf}

\subsection{Test of ``passive flame'' approaches}
We examine if various passive flame approaches can capture the nonlinear oscillation frequency and the structure of the global mode. The linear analysis is carried out on the reacting mean flow at $\mathrm{Re}=1250$ and $\phi=0.5$ by using four different linearised equations given in table \ref{tab:passive}, labelled as methods A to D. Method A, with the complete set of reacting flow equations as used throughout this study, serves as the reference calculation. Method B is the same as used by \cite{CASEL2022111695}, where the chemical heat release in the energy equation and the transport equation of $\mathrm{CH_4}$ are removed. Methods C and D use the incompressible Navier--Stokes equations, and thus exclude all fluctuations of density, species and enthalpy. The kinematic viscosity $\mu$ is uniform, chosen as the cold value $\mu(T= 300 \mathrm{K})$ and the hot value $\mu(T= 1350 \mathrm{K})$, respectively. 

\begin{table}
	\caption{Summary of active flame approach (A) and passive flame approaches (B-D) results, conducted on the mean flow at $\mathrm{Re}=1250$. The measured oscillation frequency by timestepping is 71.4 Hz.}
	\begin{center}
		\begin{tabular}{c c c c} 
			\hline
			Method & Input fields & Linearized equation & Leading eigenmodes [$\mathrm{s^{-1}}$]\\ 
			\hline	
			A& $u$, $\rho$, $Y_\mathrm{CH_4}$
			& Reacting flow & 0.2+72.0i\\ 
			\hline	
			B& $u$, $\rho$
			& Low Mach without reaction \citep{CASEL2022111695} & -5.8+66.8i\\
			\hline	
			C& $u$
			& Incompressible (cold $\mu$) & (1) 22.3+68.9i; (2) 9.8+76.9i\\
			\hline	
			D& $u$
			& Incompressible (hot $\mu$) & (1) 22.2+68.3i; (2) 9.6+74.8i\\
		\end{tabular}
	\end{center}
	
	\label{tab:passive}
\end{table}

\begin{figure} 
	\centering\includegraphics[width=0.8\textwidth]{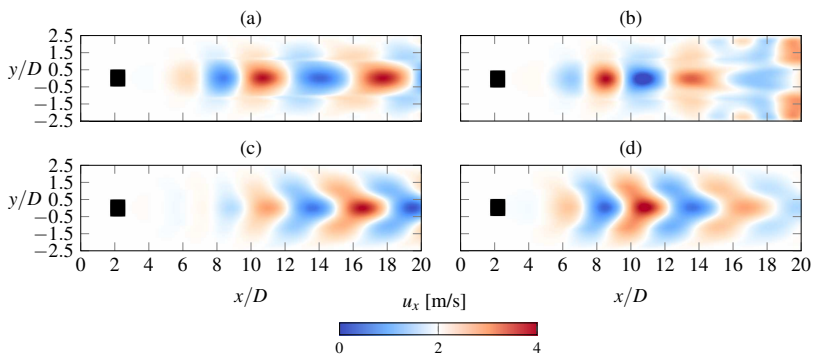}
	\caption{Leading eigenmode structures of cross-stream velocity, corresponding to methods A, B and D in table \ref{tab:passive}.}
	\label{fig:para_passive}
\end{figure}

According to table \ref{tab:passive}, method B can give reasonable predictions of the global oscillation frequency, but fails for the growth rate. Methods C and D differ qualitatively from the full model, by giving two unstable modes. The leading eigenmode structures of cross-stream velocity are presented in figure \ref{fig:para_passive}. The differences between the results of methods C and D are small, therefore only the mode structure associated with method D is shown. The general behaviour of vortex shedding is captured by all methods, yet the detailed structures along the flame fronts cannot be captured by using only the velocity as inputs through method D, as shown in figures \ref{fig:para_passive}\textit{c,d}. Including the density information into the base flow results in a slightly improved representation of flame front perturbations, shown in figure \ref{fig:para_passive}\textit{b}, but the mode structures further downstream are no longer well predicted.

%\centering\includegraphics[width=0.8\textwidth]{figures/para_kedia/para_passive-figure0.pdf}

It is thus found that the inclusion of density, species and enthalpy fluctuations into the linear model, along with their governing equations, is essential for the correct representation of the instability dynamics. This finding may be seen as conflicting with the earlier characterisation of the instability as being hydrodynamically driven, in the sense that it is underpinned by momentum fluctuation dynamics.
It must, however, be realised that the momentum fluctuations contained in an eigenmode may themselves be strongly affected by other flow quantities, in particular by density fluctuations due to unsteady reaction. The distinction between ``hydrodynamic'' and ``reaction-driven'' effects in flames has no rigorous criterion, and must always be justified in the context of the discussion.

\section{Conclusions}
\label{sec:conclusions}
This investigation characterises self-sustained oscillations of a confined cylinder-anchored flame, described by the reacting flow equations with a simple one-step model for lean methane combustion, as being the result of a linear instability. Several linear analysis tools have been leveraged to successfully predict the critical parameters for the onset of nonlinear oscillations, as well as their frequency in supercritical conditions. Wavemaker analysis has been used to infer the driving mechanisms, leading to interpretations of the stabilising effect of flame ignition and of the relative success of reduced linear models where reaction is neglected.

Performing linear eigenmode analysis on ignited and non-ignited base flows, at identical flow parameters, we first addressed the question of why ignition stabilises the flame, as has been often observed in previous studies \citep{bill1986effect,mehta2003combustion,erickson2011influence,KEDIA20142327,oberleithner2015formation,GEIKIE202128}. It is found that reaction strongly weakens the base flow shear in the portion of the recirculation bubble where the wavemaker is located. As the hydrodynamic instability in the non-ignited flow thrives on this shear, its stabilisation in the ignited flow is attributed to the base flow alteration.

With an equivalence ratio $\phi=0.5$, the ignited flow again becomes linearly unstable at a Reynolds number above 1150, in line with nonlinear time-resolved simulation that shows self-sustained oscillations setting in as this threshold is crossed. Similarly, as in classical literature on non-reacting wake flows, it is demonstrated that the leading linear eigenmode, characterised by a zero growth rate, accurately matches the frequency of the nonlinear limit-cycle oscillations, under the condition that the equations are linearised around the mean flow obtained from the nonlinear simulations. The present work represents one of the first attempts to include a linearised chemical scheme in the analysis of time-averaged reacting flows, and the results are in excellent agreement with the RZIF criterion \citep{turton2015prediction,bengana2021frequency}. Modal mean-flow instability analysis therefore holds promise for the investigation of flame instability phenomena, also in more complex configurations.

The wavemaker analysis of leading eigenmodes conducted here, pertaining to both the steady-state base flow and the time-averaged mean flow, characterises the instability as arising from momentum dynamics, and therefore being of hydrodynamic origin. This observation may be mistaken for evidence that accurate linear models may be built from the momentum and continuity equations alone, as has been done previously; however, it is demonstrated here that the inclusion of density, species and enthalpy fluctuations in the linear system is required for quantitatively accurate modelling.

\section*{Appendix: Linear analysis of a non-reacting wake}
\begin{figure}
	\centering\includegraphics[width=0.8\textwidth]{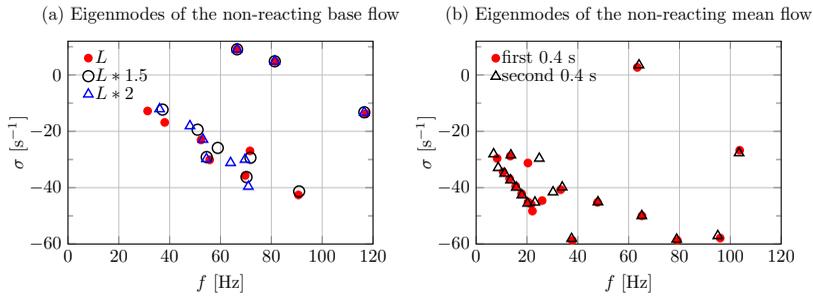}
	\caption{(a) Eigenmodes of the non-reacting base flow. The effect of domain length $L$ is tested, extending the domain downstream to $1.5L$ and $2L$. (b) Eigenmodes of the non-reacting mean flow. The effect of averaging flow is tested, with two mean flows averaged during the first 0.4 s (0-0.4 s) and the second 0.4 s (0.4-0.8 s) in figure~\ref{fig:xy_non_reacting_mean}, respectively.}
	\label{fig:xy_non_reacting_eig}
\end{figure}

Both base flow and mean flow eigenmode analysis are conducted for the non-reacting case. The steady base flow, not shown here, closely resembles the initial symmetric state in figure 14 of \citet{davis1984numerical}, whereas the mean flow is characterised by a considerably shorter recirculation zone. The eigenspectra of base and mean flow are shown in figure \ref{fig:xy_non_reacting_eig}. In both cases, the imaginary part of the leading eigenmode is close to the measured vortex shedding frequency, the mean flow yielding a more accurate result. The growth rate obtained from the base state is larger than that found from the mean state, as expected. Note that the base flow spectrum in figure \ref{fig:xy_non_reacting_eig}\textit{a} contains two unstable eigenvalues. The one with smaller growth rate has no corresponding trace in the frequency spectrum of the nonlinear signal, shown in figure \ref{fig:xy_non_reacting_mean}b.

A test of convergence with respect to the domain length is conducted by calculating the base flow with extending the flow domain downstream to $1.5L$ and $2L$ of the standard length $L$. The base state eigenspectra are given in figure \ref{fig:xy_non_reacting_eig}\textit{a}, and it is found that the least stable eigenmodes are not sensitive with respect to domain length. A convergence test is also performed for the time horizon for averaging the mean flow. Two mean flows are obtained successively by averaging over the first 0.4 s (0-0.4 s) and the second 0.4 s (0.4-0.8 s) in figure \ref{fig:xy_non_reacting_mean}\textit{a}. It is shown that the least stable eigenmodes from both mean states are well matched.  

%\centering\includegraphics[width=0.8\textwidth]{figures/non_reacting_mean/xy_non_reacting_appendix-figure0.pdf}

\section*{Declaration of Interests} 
The authors report no conflict of interest.

\section*{Acknowledgements}
The authors are grateful to Chuhan Wang's PhD jury members Laurent Gicquel, Matthew Juniper, Tim Lieuwen, Franck Nicoud, Taraneh Sayadi and Denis Sipp for their suggestions for improving this work. Chuhan Wang was supported through a PhD scholarship from Ecole Polytechnique, and by the Shuimu postdoc fellowship from Tsinghua University. Kilian Oberleithner gratefully acknowledges the Deutsche Forschungsgemeinschaft (DFG) for funding this work within the project 441269395.

\bibliography{flame.bib}
\bibliographystyle{jfm}

\end{document}